# Unravelling the spin dynamics of Molecular Nanomagnets with four-dimensional inelastic neutron scattering

Elena Garlatti,[a,b] Alessandro Chiesa,[a] Tatiana Guidi,[b] Giuseppe Amoretti,[a] Paolo Santini,[a] and Stefano Carretta*[a]

[a] Dipartimento di Science Matematiche, Fisiche e Informatiche, Università di Parma, Parco Area delle Scienze n.7/A, 43124 Parma, Italy
E-mail: stefano.carretta@unipr.it
[b] ISIS Neutron and Muon Source, Rutherford Appleton Laboratory, OX11 0QX Didcot, UK

**Abstract:** Molecular Nanomagnets have attracted the attention of the scientific community since the rich physics behind their magnetic behaviour make them ideal test-beds for fundamental concepts in quantum mechanics. Sophisticated experiments and targeted research activities have also unveiled their potential for several technological applications. Inelastic neutron scattering is a powerful and widely used technique to investigate the properties of these systems. The new generation of spectrometers, equipped with arrays of position-sensitive detectors, enable to efficiently measure the neutron cross-sections as a function of energy and of the three component of the momentum transfer vector **Q**, in vast portions of the reciprocal space. Exploiting these capabilities together with the availability of sufficiently large single-crystal samples of MNMs, it is now possible to obtain an unprecedented insight into the coherent spin dynamics of these molecular clusters. This is witnessed by several recent results, that we present in this review. By using the benchmark system $Cr_8$, it has been demonstrated that the richness of the four-dimensional inelastic neutrons scattering technique enables to extract dynamical correlation functions directly from the data. This technique has been also applied to the archetypical single-molecule magnet $Mn_{12}$ to unambiguously characterise its Spin Hamiltonian as well as to portray the entanglement between molecular qubits in $(Cr_7Ni)_2$.

## 1. Introduction

Molecular nanomagnets (MNMs) are clusters containing a finite number of paramagnetic centres, typically 3$d$ ions, whose spins are strongly coupled by Heisenberg exchange interactions.[1,2] Magnetic cores of adjacent molecules are well separated from each other by organic ligands, which make inter-molecular interactions negligible. Thus, the crystals behave like an ensemble of non-interacting identical molecules, enabling single-molecule properties to be accessed by bulk measurements. MNMs represent therefore ideal test beds for fundamental concepts in quantum mechanics such as quantum-tunnelling of the magnetization,[3,4] Berry-phase interference,[5] frustration,[6-12] finite-size effects in open or closed rings,[13-15] Néel-vector tunnelling,[16,17] entanglement[18-21] or coherence.[22-24] Moreover they are also promising building blocks for technological applications ranging from high-density data storage[25-27] to magnetocaloric refrigeration,[28-31] spintronics[32-34] and quantum information processing.[35-44] MNMs can be chemically engineered, enabling a precise control of the structure of the cores and hence of the magnetic interactions. Moreover, they can also be grafted onto surfaces keeping their properties.[45-47] From a theoretical point of view, MNMs can be accurately described by Spin Hamiltonian models,[1] that can be often simulated exactly with analytical or numerical methods.

On the experimental side, inelastic neutron scattering (INS) is the technique of choice to characterise MNMs.[48] One of its main advantages with respect to other spectroscopic techniques is that the energy spectrum of molecular clusters can be obtained without the application of any external magnetic field. Pioneering applications of INS to MNMs enabled a precise determination of the Giant-Spin Hamiltonian of the archetypal Single-molecule Magnets $Fe_8$[49] and $Mn_{12}$.[50] Besides intra-multiplet transitions due to anisotropy-induced energy splittings, INS allows one to observe transitions between different total-spin multiplets and to detect S-mixing effects leading to the breakdown of the Giant-Spin model.[51, 52] For instance, antiferromagnetic (AF) rings have been fully characterised by this technique,[13, 22, 28, 53, 54] together with many other MNMs.[55-61] Moreover, the rich physics displayed by MNMs can be probed by INS, unravelling some of the above mentioned peculiar quantum phenomena, such as Néel-vector tunnelling,[16, 17] frustration,[8, 10, 11, 62-65] and finite-size effects.[14, 66] Finally, INS has been employed to evidence the effect of pressure on the $Mn_{12}$ anisotropy barrier[67] and to monitor its relaxation dynamics.[68]

While the INS energy spectra directly yield the eigenvalues of the Spin Hamiltonian, the dependence of the scattered intensity on the transferred momentum **Q** (see Figure 1a) contains information about the eigenvectors, which is often crucial for the spectroscopic assignment of the observed excitation.[48] In particular, the dependence on the magnitude of **Q**, as obtained with powder samples, allows one to distinguish magnetic excitations from vibrational ones, as well as to discriminate between intra- and inter-multiplet transitions and to identify the symmetry of the involved eigenstates.[54, 69, 70] However, powder measurements only yield limited and partially integrated information. New possibilities in INS experiments on MNMs have been opened by the availability of sufficiently large single-crystal samples and by exploiting the new generation of high-flux cold-neutron time-of-flight spectrometers[71, 72] with large arrays of position-sensitive detectors. This, together with the advances in software,[73] enables the determination of the four-dimensional scattering function $S(\mathbf{Q}, \omega)$ over a vast portion of the reciprocal space.[74]

The aim of this review is to show the capabilities of the four-dimension inelastic neutron scattering technique (4D-INS, see Figure 1b) in unravelling the spin dynamics of MNMs. Indeed, the non-dispersive nature of magnetic excitations in these molecular clusters, i.e. the fact that transition energies are independent from **Q**, gives direct access to the spatial form of the eigenvectors, and thus makes this technique particularly powerful. Compared to its two-dimensional counterpart, 4D-INS provides a much more selective characterization of these systems, removing ambiguities between different candidate models which can be discriminated only by the vectorial $\mathbf{Q}$-dependence of $S(\mathbf{Q}, \omega)$.[75] In addition, the amount of available information is so large that the full pattern of real-space dynamical two-spin correlations can be determined, without using any model Hamiltonian.[74] Conversely, dynamical spin correlations cannot be extracted directly from powder INS measurements, because in the resulting angular-averaged cross-section formula[54] all the correlation coefficients in the three different directions x,y,z in space multiply the same $\mathbf{Q}$-dependent function and therefore cannot be extracted separately. The available information is limited even in the isotropic case, since only scalar distances between magnetic ions enter the powder cross-section.

The review is organised as follows: we first introduce (Section 2) the theoretical background, i.e. the spin Hamiltonian formalism used to model MNMs (2.1) and the basics of the 4D-INS technique (2.2), focusing on the key ingredients that can be experimentally obtained from the inelastic neutron cross-section, i.e. dynamical spin-spin correlation functions. As a benchmark case, we overview in Section 3 its application to the prototypical $Cr_8$ system. Then we present two case studies which highlight the power of 4D-INS: in Section 4 we show how the technique was employed to unambiguously determine the spin Hamiltonian of the archetypal single-molecule magnet $Mn_{12}$ (4.1) and of the highly anisotropic $Cr_7Co$ AF ring (4.2), while in Section 5 we report its use to portray entanglement between molecular qubits.

## 2. Theoretical Background

### 2.1. Spin Hamiltonian description of MNMs

All the systems presented in this review are MNMs containing $3d$ paramagnetic ions. The spin Hamiltonian (SH) describing this class of molecules usually contains the following contributions:

$$\mathcal{H} = \mathcal{H}_{iso} + \mathcal{H}_{AN} + \mathcal{H}_{ZFS} + \mathcal{H}_{Zeem}. \quad (1)$$

The first term represents the isotropic Heisenberg exchange coupling, which is the dominant one for $3d$ ions, while the second models the anisotropic contribution to exchange interaction. They can respectively be written as:

$$\mathcal{H}_{iso} = \sum_{i>j} J_{ij}\, \mathbf{s}_i \cdot \mathbf{s}_j, \quad (2)$$

$$\mathcal{H}_{AN} = \sum_{i>j} \sum_{\alpha,\beta} D_{ij}^{\alpha\beta}\, s_{i,\alpha} s_{j,\beta}, \quad (3)$$

where $\mathbf{s}_i$ are the spin operators of the $i^{th}$ magnetic ion in the molecule, $J_{ij}$ are the isotropic exchange parameters and $D_{ij}^{\alpha\beta}$ are the components of the anisotropic exchange tensors $\mathbf{D_{ij}}$. The third term of Eq. (1) represents the zero-field splitting (ZFS) interaction, arising from the combined effect of the local crystal electric fields, produced by surrounding ligand charges, and of spin-orbit coupling. The widely-used second order ZFS Hamiltonian, corresponding to a crystal field with rhombic symmetry, is written as:

$$\mathcal{H}_{ZFS} = \sum_{i=1}^{N} \left[ d_i s_{i,z_i}^2 + e_i \left( s_{i,x_i}^2 - s_{i,y_i}^2 \right) \right], \quad (4)$$

where parameters $d_i$ and $e_i$ are respectively the *axial* and *rhombic* ZFS parameters referred to the local principal axes. The last term of the SH in Eq. 1 is the Zeeman interaction with an external magnetic field:

$$\mathcal{H}_{Zeem} = \mu_B \sum_{i=1}^{N} \mathbf{B} \cdot \mathbf{g_i} \cdot \mathbf{s_i}, \quad (5)$$

where $\mu_B$ is the Bohr magneton, $\mathbf{g_i}$ is the spectroscopic splitting tensor for each magnetic ion in the molecule and $\mathbf{B}$ the external magnetic field.

All the terms of the SH can be rewritten as irreducible tensor operators and the calculation of their matrix elements can be greatly simplified by using the total-spin basis and the Wigner-Eckart theorem.[76]

### 2.2. The 4D-INS Technique

The spin-only magnetic scattering function is:[77]

$$S(\mathbf{Q}, \omega) \propto \sum_{\alpha,\beta=x,y,z} \left( \delta_{\alpha,\beta} - \frac{Q_\alpha Q_\beta}{Q^2} \right) \sum_{m,n,i,j} p_n F_i(Q) F_j(Q)$$

$$e^{i\mathbf{Q}\cdot\mathbf{R}_{i,j}}\langle n|s_{i,\alpha}|m\rangle\langle m|s_{j,\beta}|n\rangle\delta\left(\omega-\frac{E_m-E_n}{\hbar}\right), \quad (6)$$

where $F_i(Q)$ is the magnetic form factor for the $i^{th}$ ion, $|m\rangle$ and $|n\rangle$ are SH eigenstates, $E_m$ and $E_n$ are eigenenergies, $\mathbf{R}_{i,j}$ are the relative positions of the magnetic ions (determined from the known molecular structure) and $p_n$ is the thermal population of state $|n\rangle$. The term $(\delta_{\alpha,\beta} - Q_\alpha Q_\beta/Q^2)$ is a polarization factor, modeling the fact that only spin fluctuations perpendicular to $\mathbf{Q}$ couple to the moment of the neutrons. In the following we discuss experimental results obtained at very low temperature ($T \to 0$), where only the ground state is populated. Thus, in Eq. (6), $p_n = 1$ only for $n = 0$.

The spin operators $s_{i,\alpha}$ appearing in (6) can be recast in terms of rank 1 irreducible tensor operators, yielding selection rules for INS transitions, i.e. $S' - S = 0, \pm 1$ and $M' - M = 0, \pm 1$. Here $S, M$ ($S', M'$) are the total spin and its third component for state $|n\rangle$ ($|m\rangle$).

The key ingredients of 4D-INS technique are the products of spin matrix elements entering the cross-section, which are the Fourier coefficients of $T = 0$ dynamical correlation functions:

$$\langle s_{i,\alpha}(t) s_{j,\beta}(0)\rangle = \sum_m \langle 0|s_{i,\alpha}|m\rangle\langle m|s_{j,\beta}|0\rangle. \quad (7)$$

Here for simplicity we have set $E_0 = 0$. As we show in the following sections, while the values of $E_p$ are directly read out from the energies of the peaks in the INS spectrum, the Fourier coefficients $C_{ij}^{\alpha\beta} = \langle 0|s_{i,\alpha}|m\rangle\langle m|s_{j,\beta}|0\rangle$ can be extracted from the data by fitting Eq. (6) to the observed $\mathbf{Q}$-dependence of each peak.

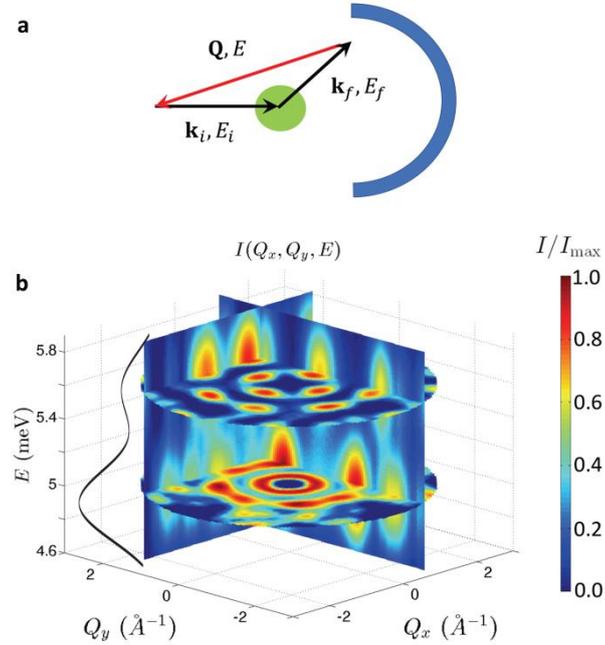

**Figure 1.** (a) Sketch of the experimental setup, with neutrons of incident momentum $\mathbf{k_i}$ and energy $E_i$ scattered after interacting with the sample to $\mathbf{k_f}$, $E_f$. The semicircle represents the array of position-sensitive detectors. (b) Four-dimensional inelastic neutron spectrum of $Mn_{12}$, as a function of the transferred energy $E = E_i - E_f$ and momentum $\mathbf{Q} = \mathbf{k_i} - \mathbf{k_f}$. Here the scattered neutron intensity is shown as a function of $Q_x$, $Q_y$ and $E$, integrated over the full measured $Q_z$ range, in the energy window containing peaks II and III reported in Figure 6 below. Adapted Figure 2-b with permission from: A. Chiesa et al., Phys. Rev. Lett. 2017, 119, 217202[82], Copyright(2017) by the American Physical Society.

## 2.3. Experimental methods

INS experiments on MNMs containing 3d ions are usually performed on cold neutron spectrometers, whose neutron incident energy range (below 25 meV) allows us to resolve magnetic transitions between different total-spin multiplets as well as to extract energy splittings of these multiplets induced by anisotropy effects. As pointed out in the previous section, crucial information is also contained in the $\mathbf{Q}$-dependence of each magnetic excitation. The *dynamical range*, i.e. the experimentally accessible ($\mathbf{Q}$, $E$) range, as well as the energy resolution, must be optimized for the particular experiment at hand. Indeed, they both depend on the incident neutron energies and scattering angles. When the incident neutron is scattered by the sample during an INS experiment it exchanges both energy $E$ and momentum $\mathbf{Q}$ with the sample, obeying the two conservation laws (see also Figure 1-a):

$$E = E_i - E_f, \quad \mathbf{Q} = \mathbf{k_i} - \mathbf{k_f}, \quad (8)$$

where the neutron has incident momentum $\mathbf{k_i}$ and energy $E_i = \hbar k_i^2/2m$ (where m is the mass of the neutron) and $\mathbf{k}_f$ and $E_f$ are the final momentum and energy, respectively. In the simple case of a scattering event in the horizontal scattering plane, if we denote the scattering angle $\varphi$ as the angle between $\mathbf{k_i}$ and $\mathbf{k}_f$, we can rewrite the momentum conservation law as:

$$k_i^2 - 2k_i k_f \cos\varphi + k_f^2 = Q^2, \quad (9)$$

and by combining it with the energy conservation law we obtain:

$$2 - 2(1 - E/E_i)^{1/2}\cos\varphi - E/E_i = Q^2/k_i^2. \quad (10)$$

Thus, after selecting incident neutrons with the desired $E_i$ and $\mathbf{k_i}$, equation (10) gives us the accessible values of the transferred energy $E$ and momentum $\mathbf{Q}$ for every scattering angle $\varphi$. The accessible ranges in both $E$ and $\mathbf{Q}$ increase if $E_i$ is increased. However, the resolution in both these variables gets worse as $\mathbf{k_i}$, and thus $E_i$, are increased.

The new generation of direct geometry cold-neutron time-of-flight (TOF) spectrometers like IN5[71] at the Institute Laue-Langevin and LET[72] at the ISIS Neutron and Muon Source are now equipped with a large array of position sensitive detectors. With direct geometry instruments it is possible to choose the energy of the incident neutrons, while measuring $E_f$ and $\mathbf{k}_f$ with the (TOF) technique combined with position-sensitive detectors. The selection of dynamical range and resolution is done by choosing the incident energy by phasing the choppers and by setting the chopper speed.. Spectrometers like LET and IN5 ensure optimal resolutions for the type of experiment discussed here. For example, the two sets of 300 Hz counter-rotating disk choppers of LET produce fine energy resolutions of $\geq 0.8\ \delta E/E_i$, while keeping an high neutron flux. The set of rotating disk-choppers in direct geometry instruments is also used for pulsing the neutron beam before the TOF analysis (for continuous neutron sources) and for the suppression of frame overalp and contamination by harmonics effects. The scattered neutrons energy and momentum are then calculated by knowing the TOF. Given the initial ($E_i$, $\mathbf{k_i}$) values and knowing the moderator-to-sample distance $D_1$ and the sample-to-detectors distance $D_2$, it is possible to determine the corresponding time-of-flights $t_1$. By measuring the time of arrival to the detector $t_2$ we can then calculate:

$$E_f = \frac{1}{2}m\left(\frac{D_2}{t_2 - t_1}\right)^2, \qquad k_f = \frac{m}{\hbar}\left(\frac{D_2}{t_2 - t_1}\right). \quad (11)$$

The presence of position-sensitive detectors enables to obtain full information on the final neutron wave-vector $\mathbf{k}_f$, an thus on the transferred momentum $\mathbf{Q}$. Indeed, the detector banks are able to identify the direction of $\mathbf{k}_f$ both in-plane and out-of-plane with respect to primary flying path, yielding a 3D information on the vectors. Given a chosen $\mathbf{k_i}$ and sample orientation, the three independent degrees of freedom are therefore the polar angles $\theta$ and $\varphi$, which define the direction of $\mathbf{k}_f$, and the time $t_2$, which is correlated to the energy transfer $E$ or equivalently to $\mathbf{k}_f$. Consequently, for a single orientation of the sample with respect to $\mathbf{k_i}$, we are able to probe a 3D volume of the four dimensional scattering function $S(\mathbf{Q}, \omega)$. The volume is defined by the accessible ranges of $\theta$ and $\varphi$ set by the size of the detectors array and by $-\infty \leq E \leq E_i$. To recover the fourth dimension, the sample is rotated about its vertical axis in small angular steps and measurements are then gathered together to build a 4D $S(\mathbf{Q}, \omega)$. Both LET and IN5 detector banks allow one to cover a wide range of the reciprocal space. LET is characterized by a π-steradiants bank divided in about $10^5$ pixels, covering 180° in the horizontal plane and ±30° out-of plane. IN5 has a 30m$^2$ position sensitive detector divided in $10^5$ pixels, covering 147° of azimuthal angle and ±20° out-of plane.

## 3. Application to the benchmark system Cr$_8$

The Cr$_8$ cluster[53, 78] was largely studied both for its intrinsic fundamental interests, and for being the precursor of the wide family of AF rings[18, 19, 54, 79]. It consists of eight Cr$^{3+}$ ions arranged on a nearly perfect octagon (inset of Figure 2). The molecule can be described by a SH with uniform magnetic couplings, whose relevant terms are only the nearest-neighbours isotropic exchange $J_{i,i+1} \equiv J$ =1.46 meV and the axial ZFS interactions $d_i \equiv d = -0.038$ meV. The anisotropy axis z is perpendicular to the ring plane.[13, 53] Cr$_8$ is characterized by a non-magnetic ground state with total spin $S = 0$ and its low-energy spectrum clearly shows rotational bands, of which the lowest two (the so-called L and E bands) approximately follow the Landé rule $E(S) = JS(S + 1)/4$ and are rigidly shifted in energy with respect to each other.[80, 81] At low temperature only the ground singlet is populated and the INS spectrum[53] shows three peaks towards $S = 1$ excited multiplets (see inset of Figure 2), associated to the transitions $p$ = 1, 2, 3. High-resolution INS measurements allow one to resolve the anisotropy splitting of the L-band $S = 1$ triplet (see below).

**3.1. Dynamical correlation functions from 4D-INS**

4D-INS experiments were performed on a large deuterated Cr$_8$ single crystal on the IN5 time-of-flight inelastic neutron spectrometer[71] at the Institute Laue-Langevin. Figure 3 shows the variation of the measured $S(\mathbf{Q}, \omega)$ in the $Q_x - Q_y$ plane for the three aforementioned transitions (panels a-c). This is characterized by several maxima and minima whose shape and positions depend on the specific transitions and reflect the structure of the involved eigenstates. By fitting these experimental maps with Eq. (6) it has been possible to extract the Fourier coefficients $C_{ij}^{\alpha\beta}(m)$. Indeed, we stress that these are the only unknown parameters in the cross-section, having determined the eigenenergies from the position of the peaks. In the present case only $C_{ij}^{\alpha\alpha}(m)$ coefficients contribute to the cross-section. These are found to be practically independent of $\alpha$, as expected for a system in which the effect of anisotropy is only perturbative. Figure 3 shows the excellent agreement between the measured $S(\mathbf{Q}, \omega)$ (panels a-c) and those calculated using the fitted values of $C_{ij}^{\alpha\alpha}(m)$ (panels d-f). These values are also very close to those obtained from exact diagonalization of the well-established SH of $Cr_8$.

The effect of anisotropy can be investigated by higher resolution measurements, allowing one to resolve the splitting of the $S = 1$ first excited multiplet into a $|M = 0\rangle$ singlet and a $|M = \pm 1\rangle$ doublet. The results are reported in the upper panels of Figure 4. The maxima have the same positions in the $Q_x - Q_y$ plane, but the relative intensities among them is different. The Fourier coefficient fitted from these maps are again in very good agreement with theory and reproduce very well the measured cross-section (panels c,d of Figure 4). It is worth noting that the obtained values of $C_{ij}^{\alpha\alpha}(m)$ reflect the axial symmetry of the SH. Indeed, in the first transition only $C_{ij}^{zz}(m)$ correlations are practically non-zero and hence is characterized by $\Delta M = 0$ (panels a, c, e of Figure 4), while in the second one (panels b, d, f) only $C_{ij}^{xx}(m) = C_{ij}^{yy}(m)$ are sizeable, indicating that $\Delta M = \pm 1$. Thus the $|M = 0\rangle$ singlet is lower in energy than the $|M = \pm 1\rangle$ doublet (easy-plane molecular anisotropy). Hence, this high-resolution experiment enable to separately determine dynamical correlations for different components of the spins.

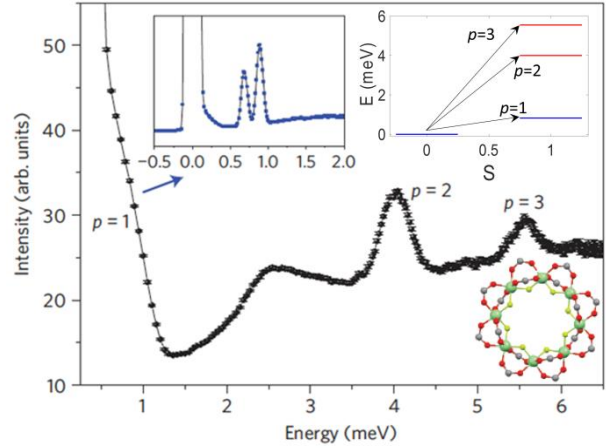

**Figure 2.** Energy dependence of the INS spectrum for a powder $Cr_8$ sample measured at low-$T$ with an incident neutron wavelength $\lambda$ = 3.1 Å. Top left inset: high-resolution powder spectrum measured with an incident neutron wavelength $\lambda$ = 5 Å, evidencing the anisotropy-induced splitting of $p$ = 1 transition. Top right: low-lying energy multiplets as a function of their total-spin S for the isotropic Hamiltonian of $Cr_8$, with arrows indicating the three transitions corresponding to the INS peaks and blue (red) lines highlighting L (E) bands. The lowest-lying S = 1 multiplet is further split by anisotropy, yielding the two $p$ = 1 transitions at low energy. Bottom left: molecular structure of $Cr_8$ ($C_{80}Cr_8F_8D_{144}O_{32}$; green, Cr; yellow, F; red, O; dark grey, C; D omitted). Adapted by permission from Springer Nature: Springer Nature, Nature Physics, "Unravelling the spin dynamics of Molecular Nanomagnets with four-dimensional inelastic neutron scattering", Michael L. Baker, *et al.*, Nat. Phys., 2012, 8, 906[74], Copyright(2012).

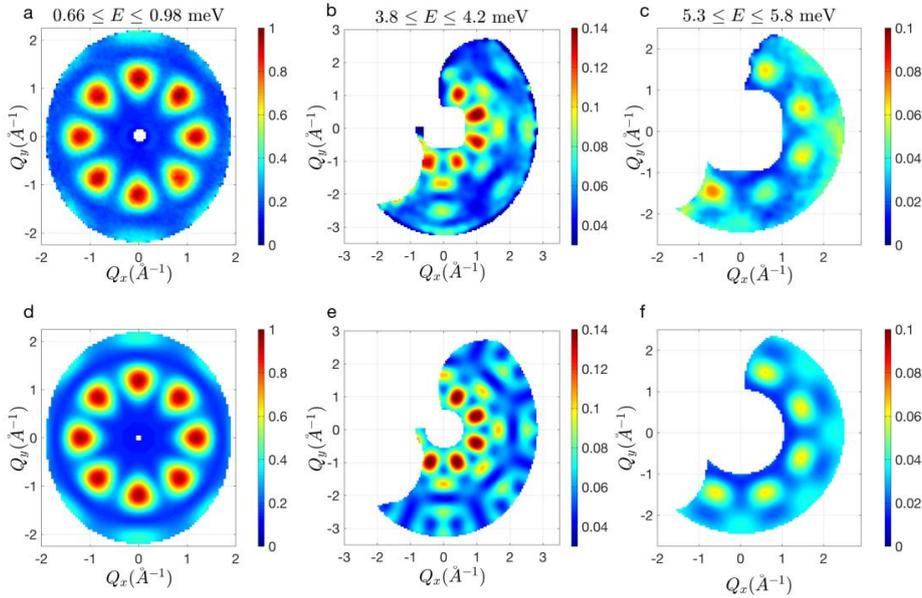

**Figure 3.** Constant-energy plots of the INS intensity for all of the possible magnetic excitations in $Cr_8$ at 1.5 K in the $Q_x - Q_y$ plane, integrated over the full $Q_z$ range. Upper panels (a-c) are the measured cross-sections for transitions 1-3, respectively. Panel a was measured with $\lambda$ = 5 Å, while panel b and c with $\lambda$ = 3.1 Å incident neutron wavelength. Lower panels (d-f) report the corresponding best fits of Eq. 6. Adapted by permission from Springer Nature: Springer Nature, Nature Physics, "Unravelling the spin dynamics of Molecular Nanomagnets with four-dimensional inelastic neutron scattering", Michael L. Baker, *et al.*, Nat. Phys., 2012, 8, 906[74], Copyright(2012).

It is important to note that the Fourier coefficients have been determined without requiring any a-priori knowledge of the system SH. Hence, the 4D-INS technique allows one to extract dynamical spin-spin correlation functions directly from experimental data. These embed crucial information on the system dynamics. Indeed, the quantities appearing in Eq. (4) are related to the response of spin $s_{j,\beta}$ at time $t$ to an instantaneous excitation applied along direction $\beta$ on spin $j$ at time $t = 0$, i.e. they can be used to visualize the real-time propagation of spin excitations along the ring.[74]

An intuitive picture on the information collected in reciprocal $(\mathbf{Q}, \omega)$ space through dynamical correlations can be obtained by translating it in terms of real-space time and position variables. This description corresponds, for each excitation, to a portray of the spin oscillations pattern of the individual spins. Indeed, the $\mathbf{Q}$ dependence of a peak at energy $E_p$ is related to the spatial pattern of the spins oscillating at frequency $E_p/\hbar$, after a resonant perturbation has brought a molecule from its ground state into a superposition state with a small component on the corresponding excited state. These patterns are determined by the same reduced matrix elements entering $S(\mathbf{Q}, E_p/\hbar)$ and thus directly descend from the associated maps. A few examples of these spin motions, directly extracted from experimental data, are depicted in Figure 4e,f for the two transitions $p = 1$ (split by anisotropy) and reflect the form of the involved eigenfunctions. In particular, the peak at 0.7 meV corresponds to a transition characterized by $\Delta M = 0$, yielding spin oscillations along $z$ axis (perpendicular to the plane of the ring, panel e). Conversely, the transition occurring at 0.9 meV is characterized by $\Delta M = \pm 1$ and hence it is associated to rotations of the spins in the x,y plane. In both cases neighbouring spins are rigidly locked in anti-parallel configuration and oscillate in phase at the frequency of the examined excitation.

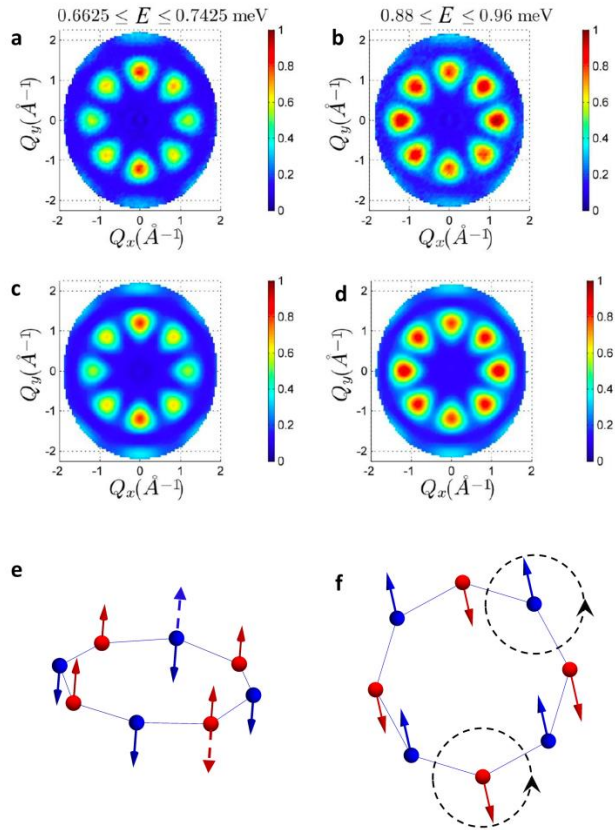

**Figure 4**. Constant-energy plots in the $Q_x - Q_y$ plane (integrated over the full $Q_z$ range) of the INS intensity for the resolved p = 1 excitation, measured at 1.5 K with $\lambda = 5$ Å incident neutron wavelength. Panels (a-b) are the measured cross-sections, while panels (c-d) are the corresponding best fits of Eq. 6. (e-f) Oscillations pattern of the individual Cr spins. Arrows represent, for each transition, the vectors $(\langle s_{i,x}(t), s_{i,y}(t)\rangle)$, while dashed circles and arrow describe the spins motion after a resonant perturbation has brought a molecule from its $S = 0$ ground state into a superposition state with a small component on the corresponding excited $S = 1, M = 0$ (left) or $S = 1, M = 1$ state (right). All the spins oscillate with the same frequency $E_p/\hbar$. Adapted by permission from Springer Nature: Springer Nature, Nature Physics, "Unravelling the spin dynamics of Molecular Nanomagnets with four-dimensional inelastic neutron scattering", Michael L. Baker, *et al.*, Nat. Phys., 2012, 8, 906[74], Copyright(2012).

## 4. Unambiguous determination of the Spin Hamiltonian

The great amount of information contained in the four-dimensional inelastic neutron cross-section can also be exploited to unambiguously determine the parameters of the system spin Hamiltonian. Indeed, in spite of impressive chemical advances in the synthesis of large and tailor-made polycentric molecules, our understanding of these systems is still limited by the difficulty in determining the interactions within the magnetic core.

### 4.1. Fingerprints of $Mn_{12}$: the forefather of MNMs

As a paradigmatic example, we illustrate the case of $Mn_{12}$-$^tBuAc$ polymetallic complex[82], belonging to the $Mn_{12}$ family of single-molecule magnets. The parent $Mn_{12}$-Ac compound can be considered the forefather of all MNMs, since the possibility to store magnetic information in a single molecule was first reported for this cluster[25]. Indeed, $Mn_{12}$-Ac gave rise to the term "single-molecule magnet", introduced to describe its magnetic bistability, and to an entirely new field of research.[3, 83] The discovery that crystals of $Mn_{12}$-Ac (crystallizing in tetragonal space group $I\bar{4}$) in reality contained several isomers breaking the true $S_4$ site-symmetries of the $I\bar{4}$ space group[84-86] lead to the synthesis of other truly high-symmetry derivatives, such as $Mn_{12}$-$^tBuAc$.[87] The breaking of S4 symmetry leads to sizable changes on the anisotropic terms of the effective giant spin Hamiltonian[86], thus influencing the relaxation dynamics[84]. However, the susceptibility (and hence the structure of the exchange multiplets) is essentially identical in the two complexes.[87] Indeed, the position of inter-multiplet transitions in the energy spectra of the two compounds are very close. [82,91]

We thus focus in the following on the high symmetry $Mn_{12}$-$^tBuAc$ derivative (from now on referred as $Mn_{12}$), which was found to be a source of better data when investigated by micro-SQUID[83] as well as solid-state spectroscopic methods.[87] It also represents an outstanding example of the aforementioned difficulty in accessing the many parameters characterizing its microscopic SH.

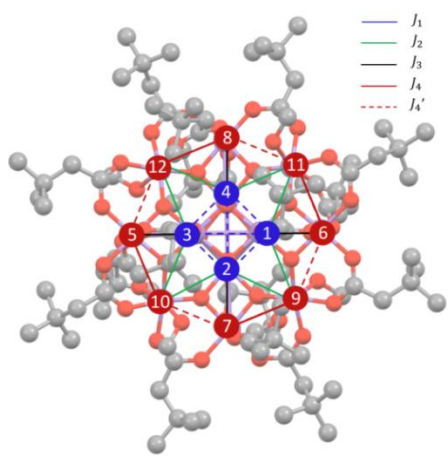

**Figure 5**: Structure of the Mn12 cluster (full formula [$Mn_{12}O_{12}(O_2CCD_2C(CD_3)_3)_{16}(CD_3OD)_4$]($C_2H_5)_2O$ which crystallizes with $S_4$ symmetry. $Mn^{4+}$ ($Mn^{3+}$) ions are represented by blue (red) circles, with solid lines of different colors indicating the relevant exchange interactions. Adapted Figure 1-a with permission from: A. Chiesa *et al.*, Phys. Rev. Lett. 2017, 119, 217202[82], Copyright(2017) by the American Physical Society.

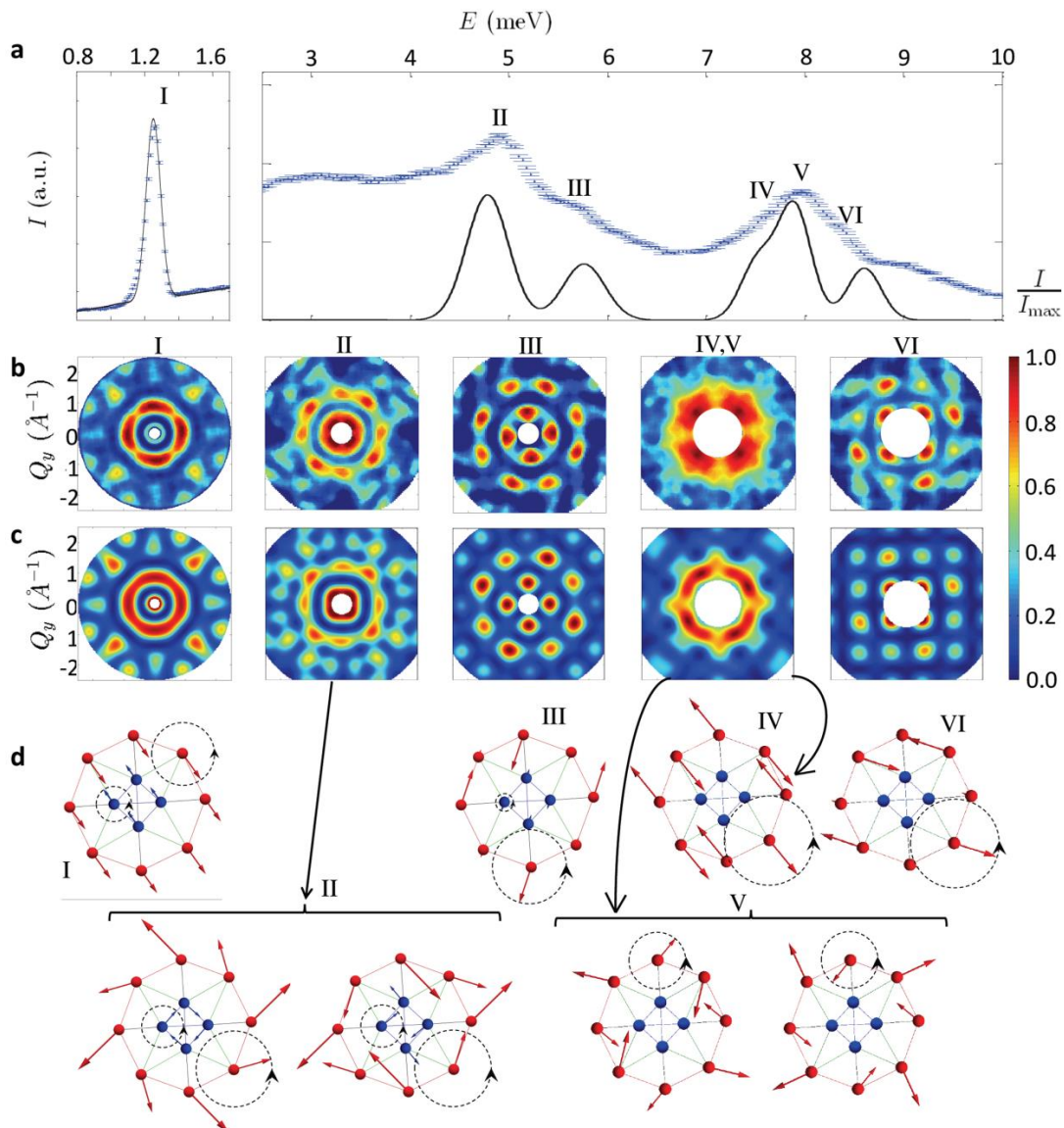

**Figure 6**. Measured energy spectrum (symbols) and corresponding calculation (continuous line) with the SH parameters (in meV) extracted from the fit of 4D-INS data: $J_1$ = -1.2 (1), $J_2$ = 3.2 (2), $J_3$ = 6.6 (3), $J_4$ = 0.55 (5), $J'_4$ = 0.30 (5), $d$ = -0.315 (2). (b) Constant-energy cuts for $S(\mathbf{Q},\omega)$, integrated over the full $Q_z$ range, obtained from measurements at T =1.5 K for incident neutron energies of 4.2 meV (first column, peak I) and 15.4 meV (peaks II-VI). Each map is normalized to its maximum. Data are also integrated over energy ranges centred around the observed transition energies: 1.12 meV < E < 1.34 meV (I), 4.6 meV < E < 5.2 meV (II), 5.3 meV < E < 5.9 meV (III), 7.2 meV < E < 8.2 meV (IV,V) and 8.2 meV < E < 8.6 meV (VI) (c) Corresponding simulated maps. (d) Precession pattern of the individual Mn spins for excitations I, II and III. Arrows represent the vectors $(\langle s_{i,x}(t), s_{i,y}(t)\rangle)$ preceding around $z$, after a resonant perturbation has brought a molecule from its $M$ = 10 ground state into a superposition state with a small component on the corresponding excited $M$ = 9 state. The two panels for excitation II and V correspond to a pair of degenerate states. Notice also that it was not possible to resolve peaks IV and V, which are very close in energy. Adapted Figure 3 with permission from: A. Chiesa *et al.*, Phys. Rev. Lett. 2017, 119, 217202[82], Copyright(2017) by the American Physical Society.

Indeed, only the application of 4D-INS to single crystals of Mn$_{12}$ allowed us to univocally fix the leading exchange interactions of this archetypal molecule, after more than 25 years of research and hundreds of studies.[88, 89] Indeed, the measured **Q**-dependent maps for different transitions from the molecular ground state, by portraying the precession of individual spins when excited by incident neutrons, provide a sort of fingerprints of the molecular eigenstates.

Most of the models proposed for Mn$_{12}$ are based on the four exchange interactions reported in the schematic picture of Figure 5, with $J_4 = J'_4$. The SH also includes anisotropic zero-field splitting axial terms, inducing a splitting of the different total spin multiplets. In particular, as already demonstrated by INS,[50] magnetometry, EPR[87] and NMR[90] measurements, the ground state is an $S = 10$ split by ZFS easy-axis anisotropy. The results presented below do not critically depend on the form of these ZFS terms, which are thus not discussed here.

Previous INS measurements on powder $Mn_{12}$ samples,[91] as well as the low-$T$ single-crystal spectrum reported in Figure 6a, supply the excitation energies from the molecular ground-state. These, however, provide only a coarse characterization of the spin Hamiltonian through its eigenvalues, while 4D-INS gives access to the selective information associated with the structure of eigenvectors. 4D-INS spectra measured on LET cold-neutrons time-of-flight spectrometer at $T = 1.5$ K are shown in Figure 6b. In particular, the first panel on the left reports the **Q** dependence of the intra-multiplet transition (peak I) between the ground $|M = \pm 10\rangle$ and first excited $|M = \pm 9\rangle$ doublet. By fitting the neutron cross-section, we can directly extract[82] from the pattern of **Q** modulations (even without any SH model) the distribution of the giant-spin moments over different Mn ions, i.e., the set of expectation values $\langle S = 10, M = \pm 10|s_{i,z}|S = 10, M = \pm 10\rangle$, which are in line with previous NMR studies[90]. The moment distribution reveals ferromagnetic correlations among the four internal $Mn^{4+}$ and among the eight external $Mn^{3+}$ spins, with the two sets antiferromagnetically correlated to each other, but with moments significantly below saturation, indicating that the spins are not locked in a maximally aligned state due to quantum fluctuations. While this moments distribution is stable over a wide set of exchange parameters, the **Q**-dependence of the other cold transitions (reported in Figure 6) is much more selective and allowed us, by an extensive exploration of the parameters space (combined also with susceptibility measurements) to definitely fix the exchange interactions of $Mn_{12}$, reported in the caption of Figure 6. The agreement between measured (b) and calculated (c) $Q_x - Q_y$ maps (obtained by diagonalizing the SH with the fitted exchange constants) is apparent.

Spin precession patterns analogous to those extracted on Cr8 are depicted in Figure 6-(d). At difference from the motions reported in Figure 4 (showing anti-ferromagnetic correlations and an increase of the total spin from the ground singlet to the excited triplet) here they correspond to the molecular counterpart of spin-wave excitations in bulk ferromagnets. The different spin dynamics associated to the various transitions is evident: in transition I all the spins rigidly precess conserving the same total-spin modulus of the ground state, indicating that this is a giant spin excitation. Conversely, all the other peaks display a precession pattern characterized by a zero total spin, a direct demonstration of the inter-multiplet nature of the transitions.[92] Furthermore, different symmetries of the excited states produce evident signatures in the precession patterns.[82]

**4.2. The anisotropic AF ring Cr₇Co**

The Cr₇Co ring represents a model system to understand how the insertion of a highly anisotropic $Co^{2+}$ ion into an AF ring determines the anisotropy of the whole molecule. Indeed, it has been demonstrated by INS that the anisotropy of a $Co^{2+}$ ion is transferred to the whole polymetallic cluster through strong exchange interactions.[93] On the one hand, it is of fundamental importance to deeply understand magnetic exchange interactions involving anisotropic ions. The combination of a sizeable anisotropy with exchange interactions can be a crucial point in the design of new anisotropic MNMs for applications in, for example, quantum information processing.[40, 94] Moreover, this kind of compounds is expected to display rich physics associated with the unquenched orbital degrees of freedom, e.g. Néel-vector tunneling.[16, 17, 68]

Cr₇Co molecular structure is similar to that of Cr₈, with magnetic ions at the vertexes of an octagon. It contains seven isotropic $Cr^{3+}$ ions, behaving as pure spin $s_{Cr} = 3/2$, and one anisotropic $Co^{2+}$ ion, that can be described at low temperatures as an effective spin $s_{Cr} = 1/2$, with highly anisotropic **g** tensor and exchange interactions.[95] In order to survey Cr₇Co SH and in particular to investigate its anisotropy, several INS measurements have been performed, both on powder and single crystal samples and also in applied magnetic field. Single- crystal measurements performed at low temperature on IN5 spectrometer enabled to obtain one crucial information about the effect of anisotropy on Cr₇Co energy spectrum, by detecting a low-energy transition below 0.1 meV (see the shoulder in Figure 7). In these single-crystal measurements the 4D-INS technique enabled to check the magnetic origin of the 0.1 meV peak by extracting the dependence of its INS intensity on $Q_x - Q_y$. This is reported, together with the analogous map for the 0.5 meV peak, in the insets of Figure 7. The modulations in these maps clearly identify magnetic transitions. These measurements, combined with ab-initio calculations based on DFT+MB approach[96,97] and in-field INS experiments, allowed to demonstrate that these two peaks correspond to transitions within the lowest $S = 1$ multiplet. This result indicates the presence of strong isotropic ($J_{Cr-Co} = 19$ K) and anisotropic Cr-Co exchange interactions. Indeed, the rhombic anisotropic exchange tensor **D** yields two directions where the Cr-Co anisotropic exchange is strong but opposite in sign (of the order of 10 K) and a third direction where it is almost one order of magnitude weaker.[93] With these parameters, our SH model reproduces all the set of experimental data, including the intensity maps reported in Figure 7. These results demonstrate a strong and highly anisotropic exchange interaction between $Co^{2+}$ and the neighbouring Cr ions, which effectively transmits the anisotropy of $Co^{2+}$ to the whole molecule. Thus, Cr₇Co also represents a starting point for the design of new systems for quantum information processing applications, where high-spin ions are strongly coupled to a few very-anisotropic ions like $Co^{2+}$.

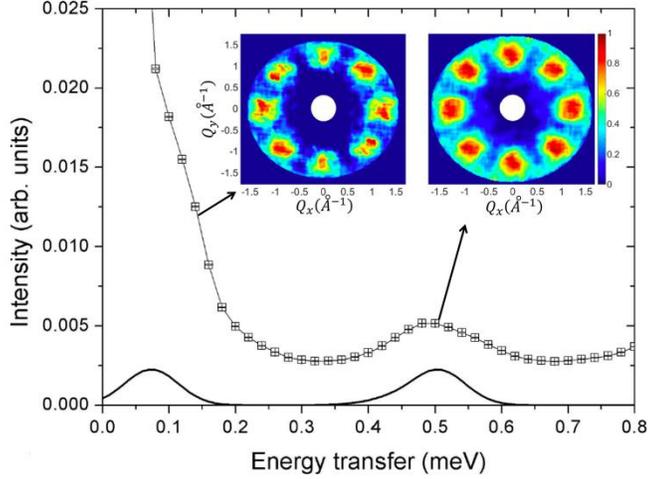

**Figure 7.** INS spectrum for a Cr$_7$Co single crystal collected on the IN5 spectrometer at T = 1.5 K with an incident neutron wavelength $\lambda$ = 6.5 Å and integrated over the full measured **Q** range (black squares). The c axis of the single crystal is perpendicular to the horizontal scattering plane. The solid line is the simulation obtained with the SH best fit including strong isotropic and anisotropic Cr-Co exchange interactions.[93] Insets: constant-energy plots of the Cr$_7$Co INS transitions at 0.1 and 0.5 meV at $T$ = 1.5 K. The measured maps in the $Q_x - Q_y$ plane are integrated over the full experimental $Q_z$ range, with an incident neutron wavelength $\lambda$ = 6.5 Å. Data are also integrated over energy ranges centred around the observed transition energies: 0.07 meV < E < 0.2 meV and 0.34 meV < E < 0.66 meV respectively. The colour-bar reports the transition intensity normalized for the maximum in each panel. The maps show the same pattern of intensity modulations in the explored **Q** range and all the maxima in each map have the same intensity, demonstrating the same occupation probability for Co$^{2+}$ of all the ring sites. Indeed, the differences in the intensity pattern of each excitation are averaged-out by the delocalization of Co$^{2+}$ along the ring. Adapted from Ref. [93] with permission from the Royal Society of Chemistry.

## 5. Portraying entanglement between molecular qubit

The 4D-INS technique has also demonstrated to be enough powerful to yield a deep insight into the entanglement between molecular qubits. Entanglement is in fact one of the most intriguing aspects of quantum mechanics[98-100] and it represents an essential resource for quantum information processing applications. It occurs when a composite quantum object is described by a wave function that is not factorised into states of the objects components, making it impossible to describe a part of the system independently of the rest of it. How to detect and quantify entanglement experimentally has become a crucial step for the development of quantum information protocols. Indeed, many recent theoretical and experimental works have focused on the detection of entanglement,[101-104] but measuring entanglement in complex system is very difficult. Arrays of weakly-coupled MNMs represent an ideal playground for investigating quantum entanglement between complex spin systems. The prototype (Cr$_7$Ni)$_2$ supramolecular dimer has in fact proven to be the ideal benchmark system to demonstrate the capability of 4D-INS to detect and quantify entanglement[21].

Among all MNMs, the Cr$_7$Ni AF ring is the most studied molecular qubit.[23, 54, 79] It contains seven isotropic Cr$^{3+}$ ions ($s_i = 3/2$) and one Ni$^{2+}$ ion ($s_i = 1$). The dominant interaction in this ring is the nearest-neighbour AF exchange (Eq.2), leading to an isolated ground doublet behaving as a total spin $S = 1/2$, which can be used to encode a qubit. AF rings can also be linked together to obtain supramolecular structures fitting specific quantum information processing schemes[18, 40] and arrays of AF rings can also be used as quantum simulators of different model Hamiltonians.[35]

The [Cr$_7$NiF$_3$(C$_7$H$_{12}$NO$_5$)(O$_2$CC(CH$_3$)$_{15}$)]$_2$(N$_2$C$_4$H$_4$) supramolecular dimer (where N$_2$C$_4$H$_4$ =N-methyl-D-glucamine), in short (Cr$_7$Ni)$_2$, consists of two Cr$_7$Ni purple AF rings, and hence is a complex system made of sixteen interacting spins (see Figure 8). The two individual rings have been fully characterized by INS and EPR spectroscopy and by low-temperature specific heat and magnetometry measurements, which have allowed us to determine all the SH parameters[105]. The pyrazine unit in the (Cr$_7$Ni)$_2$ dimer links the two single rings, providing two N-donor atoms binding to Ni centres. This cause a weak exchange coupling between the two Ni ions in ring A and ring B for the dimer:

$$\mathcal{H}_{int} = j\, s(Ni_A) \cdot s(Ni_B). \quad (12)$$

The presence of a sizeable exchange interaction between the two rings leads to entangled states for the supramolecular system. At zero field (with $j > 0$) its energy spectrum is characterised by an entangled singlet ground state and by an excited triplet. An external magnetic field can be used to induce a factorised ferromagnetic ground state, as shown in the magnetic field dependence of the energy levels in Figure 9a. All the INS measurements on (Cr$_7$Ni)$_2$ have therefore been performed at $B$ = 2.5 T and $T$ = 1.2 K to place the system into the factorized reference ground state and to univocally test the occurrence of entanglement in the two first excited states. From our model, two INS transitions are expected, corresponding to excitations from the factorised ground state towards the two excited Bell states (red arrows in Figure 9a). Two inelastic peaks have indeed been observed at 0.24 meV and 0.28 meV both on LET (ISIS) and on IN5 (ILL). These findings are reproduced with and inter-ring coupling $j$ = 1.1 K.

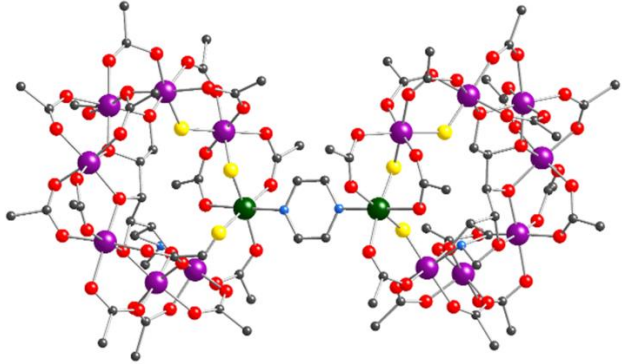

**Figure 8.** Structure of the (Cr$_7$Ni)$_2$ supramolecular dimer (magenta, Cr; green, Ni; yellow, F; red, O; light blue, N; grey, C; C(CH$_3$)$_3$ groups and all H atoms omitted for clarity). Adapted from E. Garlatti *et. al.,* Nat. Commun., 2017, 8, 14543[21], under a Creative Commons Attribution 4.0 International License.

A collection of (Cr$_7$Ni)$_2$ single crystals have been measured by 4D-INS on the IN5 spectrometer. Two conflicting requirements made this experiment particularly challenging: high-resolutions, in order to be able to resolve INS transitions within the dimers lowest-energy manifold, whose splittings are due to the small inter-ring coupling, and, at the same time, the need for a detailed **Q**-dependence of the scattering function over a large range of **Q**. Indeed, as stated in Section 2.1, the dependencies of the transition intensities on **Q** yield direct information on the dynamical spin-spin correlation functions. In particular, if we have INS transitions from a factorised reference ground state $|0\rangle$ to an entangled excited state $|m\rangle$ as in our dimer, dynamical correlations are non-zero also for pairs of spins belonging to different rings. These long-distance correlations produce short-**Q** modulations in the $S(\mathbf{Q}, \omega)$ through the exponential term $e^{i\mathbf{Q}\cdot\mathbf{R}_{i,j}}$. Hence, the pattern of maxima and minima in the measured scattering function as a function of the **Q** vector components is a sort of portrayal of the entanglement between molecular qubits in the excited state of the dimer.

Figure 9-b,c reports the measured INS intensity of the two observed transitions as a function of the $Q_x$ and $Q_z$ components of the momentum transfer vector. Short-**Q** modulations of the intensity are evident for both transitions, with the maxima in the low-energy excitation corresponding to minima in the high-energy one and vice-versa. These modulations would be absent if correlations between the two rings were forced to zero in Eq. 12. Thus, 4D-INS data clearly demonstrate the occurrence of entanglement in the excited states of the supramolecular dimer and they coincide with the results predicted by the theoretical model (see Figure 9d,e).

### 5.1. Quantification of entanglement with 4D-INS

From the richness of 4D-INS data it is possible not only to experimentally show the occurrence of entanglement, but also to extract the concurrence $C$ of the two excited states of the dimer. Concurrence $C$ is a largely used measure of entanglement between a pair of qubits, whose values range from 0 for factorised states to 1 for maximally entangled ones[106]. For pure two-qubit state $|m\rangle = a|00\rangle + b|10\rangle + c|01\rangle + d|11\rangle$ the concucurrence is

$$C = 2|ad - bc| . \qquad (13)$$

In the case of a molecular qubit, the $|0\rangle$ and $|1\rangle$ states are encoded in the two lowest-energy eigenstates, like the ground $S = 1/2$ doublet of Cr$_7$Ni. In the supramolecular dimer (Cr$_7$Ni)$_2$, both molecular qubits A and B are in this computational basis and we observed transitions from a factorised ground state $|00\rangle$ to excited $|m\rangle$ states. Thus, the concurrence of the excited states only depends on the b and c projection coefficients, since a = 0 because ⟨00|m⟩ = 0, and we have $C = 2|bc|$. The scattering function for these transitions can be re-written as a sum of two single-qubit contributions, plus an interference term[21]:

$$S(\mathbf{Q}, \omega) = |b|^2 I_{AA}(\mathbf{Q}) + |c|^2 I_{BB}(\mathbf{Q}) + \quad (14)$$

$$+2Re[bc^* I_{AB}(\mathbf{Q})],$$

where INS intensities $I_{AA}(\mathbf{Q})$, $I_{BB}(\mathbf{Q})$ and $I_{AB}(\mathbf{Q})$ are written in terms of effective spin-1/2 operators acting in the ground doublet of each molecular qubit. These quantities are also already known for Cr7Ni, as all their ingredients can be preliminary determined from measurements on single-qubit compounds.[14, 107, 108]

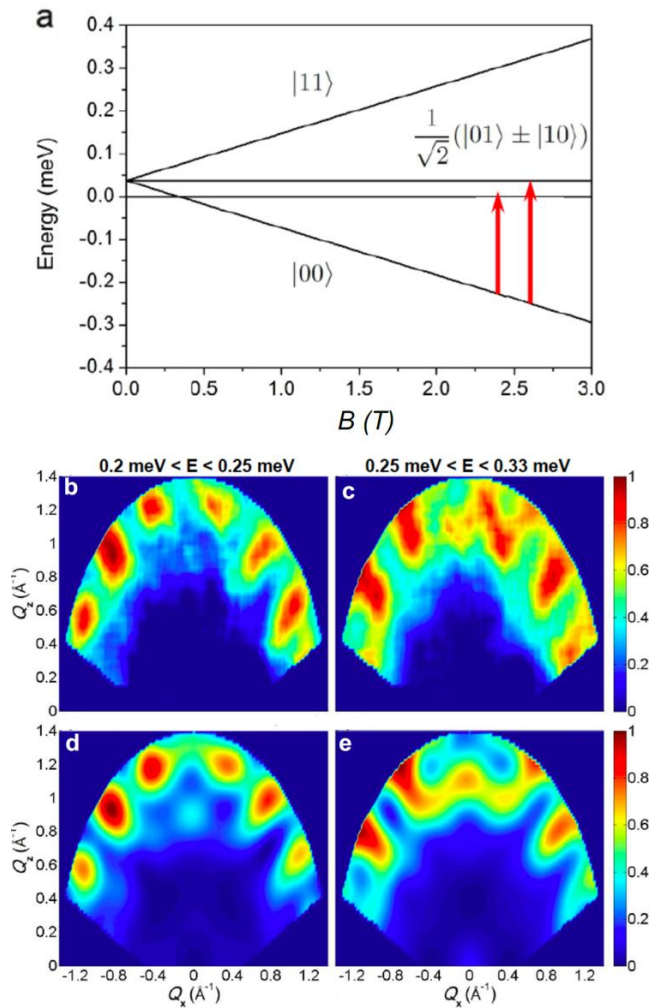

**Figure 9**. (a) Magnetic field dependence of the lowest-energy levels of the supramolecular dimer. Red arrows indicate the two INS transitions. |0⟩ and |1⟩ qubit states are encoded in the ground $|S = 1/2, M = \pm 1/2\rangle$ doublet of each ring. In applied magnetic field $B > 3.4$ T the dimer is in a factorized ferromagnetic state and the two first excited states are entangled Bell states. (b-e) Constant-energy plots of the (Cr7Ni)2 INS transitions at 0.24 and 0.28 meV. Panel b and c report the measured maps in the $Q_x - Q_z$ plane integrated over the full experimental $Q_y$ range, with an incident neutron wavelength $\lambda = 7.5$ Å Measurements were performed on IN5 with a sample temperature of 1.2 K and a magnetic field $B = 2.5$ T applied along the y-vertical axis. Panels d and e show the corresponding calculations accounting for the presence of differently oriented dimers in the crystals.[21] The colour bar reports the transition intensity normalized for the maximum in each panel. Adapted from E. Garlatti *et. al.*, Nat. Commun., 2017, 8, 14543[21], under a Creative Commons Attribution 4.0 International License.

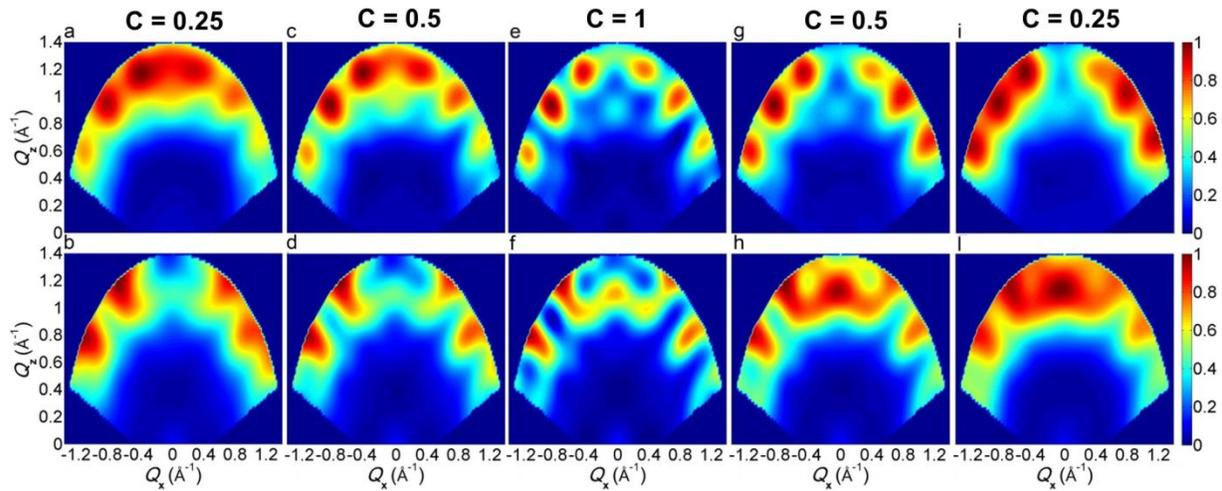

**Figure 10.** Calculated constant-energy plots of the $(Cr_7Ni)_2$ INS transitions at 0.24 (upper panels) and 0.28 meV (lower panels) for different values of the projection coefficients $b$ and $c$ of the excited states $|m\rangle$. The colour bar reports the transition intensity normalized for the maximum in each panel. Real wavefunctions with positive values of $c$ increasing (decreasing) from left to right are shown in the first (second) row and lead to different values of the concurrence $C$. Adapted from E. Garlatti *et. al.*, Nat. Commun., 2017, 8, 14543[21], under a Creative Commons Attribution 4.0 Internationall License.

Hence, by comparing experimental data with the scattering function calculated with Eq.14, we can determine the coefficients $b$ and $c$ and therefore extract the concurrence $C$ of the excited $|m\rangle$ states.

Figure 10 reports INS intensities in the $Q_x, Q_z$ plane for the two observed transitions of the dimer, calculated assuming different values of the projection coefficients $b$ and $c$, corresponding to different values of the concurrence. When the excited states are entangled and $C = 1$ (panel e and f of Figure 10) the short-**Q** modulations are clearly visible, while they tend to disappear by lowering the concurrence. By comparing our experimental data in Figure 9 with these calculations, we can state that for both excited states of the dimer $(Cr_7Ni)_2$ the concurrence $C \simeq 1$ and therefore they are maximally entangled.

The scattering function in Eq. 14 can also be generalized to other types of molecular doublets. Moreover, this whole method based on the 4D-INS technique can be applied independently from knowing the specific form of the inter-qubit interaction. Hence, entanglement can be investigated even between molecular qubits where it is not possible to obtain a sound description of the full SH.

## Summary and Outlook

In conclusion, we have reviewed the capabilities of four-dimensional inelastic neutron scattering in probing the spin dynamics of Molecular Nanomagnets. The power of this technique relies on the huge amount of information embedded in the four-dimensional cross-section collected on single-crystal samples. This requires the combination of different capabilities: on the one hand the development of new synthetic approaches to obtain large single crystals, on the other hand advances in INS instrumentation. Indeed, in the last years great progresses has been made in the synthesis of MNMs and the chemists are now able to provide high-quality single-crystals suitable for 4D-INS experiments.[21,74,82,93] Moreover, the new generation of cold neutron time-of-flight spectrometers like LET and IN5 combining high flux, low backgrounds, high resolutions and position-sensitive detectors allow one to perform this kind of experiments on relatively small samples. Finally, improvements in the flux of the spectrometers under development at the European Spallation Source (https://europeanspallationsource.se), will further expand the possibilities of this kind of experiments, opening new avenues in the understanding and design of several classes of MNMs.

In particular, there exist many polymetallic clusters that, in spite of their fundamental or applicative interest, are still little explored. We mention, among others, frustrated systems whose complex spectrum cannot be rationalized by simplified models[109] or highly anisotropic molecules containing rare-earth or Co ions which can transfer their large anisotropy to the whole cluster via strong exchange interactions.[89] In such polycentric molecules, the complexity of the spin structure and the large dimension of the associated Hilbert space often preclude their unambiguous characterization by conventional techniques. Conversely, four-dimensional Inelastic Neutron Scattering directly yields the crucial quantities necessary to describe the spin dynamics of these systems, i.e. spin-spin dynamical correlation functions. In particular, we have shown that these can be used to extract the parameters of the microscopic spin Hamiltonian, as well as to portray entanglement between molecular qubits, thus demonstrating that 4D-INS is an unrivaled technique to investigate these systems.

## Acknowledgements

The results reviewed in this paper were obtained thanks to the collaboration with many exceptional chemists and experimental physicists. The synthesis of all the compounds was performed by Prof. Richard E. P. Winpenny's outstanding team in Manchester, especially Dr. Grigore A. Timco. Further characterizations of the compounds were directed by Prof. Eric J. L. McInness. We gratefully thank Dr. Mike Baker, Dr. Jacques Ollivier and Dr. Hannu Mutka for their collaboration during INS experiments at Institute Laue-Langevin in Grenoble.

E. G., A. C., G. A., P. S. and S. C. acknowledge financial support from PRIN Project 2015 No. HYFSRT of the MIUR (Italy). E. G. and A. C. also acknowledge the support of "Fondazione Angelo della Riccia". The authors gratefully acknowledge the Science and Technology Facilities Council (STFC) for access to neutron beamtime at ISIS, and the Institute Laue-Langevin for financial support and neutron instrument time.

**Keywords:** Molecular Nanomagnets, Spin dynamics, Inelastic neutron scattering